\documentclass[11pt]{article}
\usepackage[
    natbib=true,
    style=numeric,
    sorting=none
]
{biblatex}
\usepackage[utf8]{inputenc}
\usepackage[T1]{fontenc}
\usepackage{graphicx}
\usepackage{grffile}
\usepackage{longtable}
\usepackage{wrapfig}
\usepackage{rotating}
\usepackage[normalem]{ulem}
\usepackage{amsmath}
\usepackage{textcomp}
\usepackage{amssymb}
\usepackage{capt-of}
\usepackage[bookmarks]{hyperref}
\usepackage{color}
\usepackage{listings}
\usepackage[margin=1in]{geometry}
\usepackage{authblk}
\author[1,*]{Brett Viren}
\author[1]{Jin Huang}
\author[2]{Yi Huang}
\author[2]{Meifeng Lin}
\author[2]{Yihui Ren}
\author[3]{Kazuhiro Terao}
\author[1]{Dmitrii Torbunov}
\author[1,2]{Haiwang Yu}
\affil[1]{Dept. of Physics, Brookhaven National Laboratory}
\affil[2]{Computer Science, Brookhaven National Laboratory}
\affil[3]{SLAC National Accelerator Laboratory}
\affil[*]{Corresponding author: bv@bnl.gov}
\date{\today}

\title{Solving Simulation Systematics in and with AI/ML\\\medskip \large Bridging the systematic gap between simulated and real data}

\hypersetup{
 pdfauthor={Brett Viren},
 pdftitle={Solving Simulation Systematics in and with AI/ML},
 pdfkeywords={},
 pdfsubject={},
 pdfcreator={Emacs 27.2 (Org mode 9.4.6)}, 
 pdflang={English}}
\begin{document}

\maketitle

\begin{abstract}
  Training an AI/ML system on simulated data while using that system to infer on data from real detectors introduces a systematic error which is difficult to estimate and in many analyses is simply not confronted. 
  It is crucial to minimize and to quantitatively estimate the uncertainties in such analysis and do so with a precision and accuracy that matches those that AI/ML techniques bring. 
  Here we highlight the need to confront this class of systematic error, discuss conventional ways to estimate it and describe ways to quantify and to minimize the uncertainty using methods which are themselves based on the power of AI/ML.
  We also describe methods to introduce a simulation into an AI/ML network to allow for training of its semantically meaningful parameters.
  This whitepaper is a contribution to the Computational Frontier of Snowmass21.
\end{abstract}

\pagebreak

\setcounter{tocdepth}{1}

\section{Introduction}
\label{sec:intro}

A common AI/ML application in HEP and elsewhere adopts a strategy like:

\begin{enumerate}
\item Train a network on samples from a simulated detector data domain.

\item Apply the trained network to samples from a real detector data domain.

\item ???

\item Profit.\footnote{With apologies to South Park.}
\end{enumerate}
Of course, our simulations are always wrong at some level. 
Great pains are taken to minimize this wrongness, but the result can only be as accurate as the precision of our validation methods.
We are best to keep in mind that the two data domains (simulated and real) remain distinct. 
When we fail to fill in the ``???'' with systematic error analyses of a precision commensurate with that of the primary analyses then we make what is called here a ``domain-equivalency mistake''.
With so much of our mainline analyses exploiting the high precision of artificial-intelligence and machine-learning (AI/ML) we must also evaluate our systematic errors at an equal level.
Failing that, we do not know how wrong we truly are.

Here, we discuss select ways to supply step three.
First we describe some conventional approaches to reducing the difference between the two domains.
Then we look at a special case where learned, generative AI/ML techniques are employed, give some problems with their standard validation methods and possible ways to address these problems.
Next we introduce a new concept being developed at Brookhaven National Lab for bridging the two data domains. 
And finally describe new techniques being explored at SLAC National Lab and others for integrating simulation to AI/ML training. 
Though these problem are general to many fields and applications but we discuss them in the context of recent developments and applications toward DUNE and other liquid argon time projection chamber (LArTPC) detectors.

\section{Transforms for closer equivalence}
\label{sec:closer}

Prior to the rise of machine learning, we used our ``natural intelligence'' to understand our detectors and minimize the errors made by its simulation. 
Of course, we still do this. 
We make our simulation produce data as similar to real detector data as it is feasible to do so. 
We typically strike a balance between effort and quality so the fidelity is ``good enough'' for a given purpose. 
As the high-precision of AI/ML is applied to analyses, the fidelity of the simulation must be raised to an equivalent level.

Some features of the real detector data are a ``nuisance''.
They are particularly difficult to model with required precision yet are also are not related to the physics information we seek to extract from our detectors. 
For example, real LArTPC detectors typically have some forms of coherent noise which greatly confound the extraction of signal and which are also impractical to model with high precision.
As a practical and effective work around, we apply transformations to both the simulated and the real data which attempt to minimize the prevalence of ``nuisance'' features while leaving unchanged those due to signals we wish to extract.

In the LArTPC arena, we have collected expertise in to a variety of such transforms.
We have developed software ``noise filters'' targeting a set of noise features that are specific to certain detectors or general across all.
Transforms that attack coherent noise include explicit ``signal protection'' guards so they retain certain features which can not be readily distinguished as being from signal.
We also apply crucial ``signal processing'' to amplify signal by deconvolving detector response.
As a side effect, this reduces both the general high frequency thermal noise from electronics as well as low frequency noise that would otherwise be amplified by the deconvolution.
The capability of these transforms to bring the simulated and real data domains closer has been well validated~\cite{mbsigproc1},\cite{mbsigproc2}.

However, these validations are rough in nature compared to the precision of AI/ML.
Evaluation is in terms of metrics formed as some average over the features of samples from each data domain.
The averaging hides fine features which can nonetheless still be noticed by AI/ML networks.
The ``domain-equivalency mistake'' is reduced by applying this manner of validation but how much error remains and how to measure it is still a concern.

\section{Generative network evaluations}
\label{sec:gan}

Generative adversarial networks (GANs) are class of AI/ML networks with many practical benefits.
The GAN architecture consists of two parts. 
A generator part is trained to produce samples from an underlying but ``ersatz'' distribution such they are indistinguishable from those drawn from a reference distribution.
Simultaneously, a discriminator part is trained to distinguish generated samples from reference samples.
We consider the ``ersatz'' and reference distributions distinct and samples from the latter are typically provided by a detailed, conventional simulation. 

In a scientific data context, GANs are often used to implement a ``fast simulation'' and in section~\ref{sec:ls4gan} we introduce another way to exploit them. 
The goal in this case is to replace a computationally expensive conventional simulation with a faster GAN network.

However, such a replacement introduces the potential for another domain-equivalency mistake.
The network may not fully learn the reference distribution due to use of a limited size training set and/or non-optimal network architectures.
We must keep in mind that the ``ersatz'' distribution is thus twice-removed from the one sampled by the real detector.
Next, we describe ways to get a handle on these errors.

\subsection{Network based metrics}

A GAN as a fast simulation is still useful if we can measure the quality of samples it produces. 
When the samples are in the form of 2D pixel images measures such as the Fréchet~\cite{fid} (FID) and/or kernel~\cite{kid} (KID) \emph{inception distance} (together xID) are commonly applied. 
Both measure a certain vector distance between two sets of samples.
The vectors are formed from high-level features extracted from the 2D pixel images with a help of the Inception network.
An important detail for here is that the features (labels) and the data set used to train the network are fixed by the metric definitions. 
The training uses a large, standardized dataset called ImageNet. 
It includes images of cats, cars and cardigans but certainly not any images from our detectors.

Thus, an image vector derived from a sample, be it real, simulated or ersatz, represents the result of a projection from the actual distribution domain to a domain spanned by the set of internal features associated to the ImageNet labels.
For example, one element of the vector may represent in some abstract way a ``\%-catness'' of an image. 
Applying xID as metrics for a GAN trained in scientific domains is like asking how much ``catness'' does the GAN preserve in the detector data. 
It is an absurd, if adorable, metric to judge scientific worthiness of a detector simulation.

Furthermore, xID metrics are themselves ``metric free'' (in the sense of a metric tensor).
We can not assign any absolute interpretation to its distance values across the latent space.
Furthermore, xID posses some known biases which confound attempts to compare the reported values of the ``same'' metric across different applications. 
While they are a practical tool rank different GANs trained on and compared with identical data, they are not useful for providing a quantitative systematic uncertainty measure.

The manual human effort put into the labeling of ImageNet images was certainly nontrivial. 
The similar kind of effort to label millions of detector events would exhaust even the most diligent pool of grad students. 
Happily, with simulated data our ``labels'' may be our ``Monte Carlo truth'' values. 
We may generate labeled images in massive quantities with little human effort. 
These images can then replace the ImageNet training set and a variant of the xID metrics can be constructed which is tailored to a given data domain. 
By construction such a metric will not be subject to the projection problem suffered with the xID metrics. 
Though, we must recognize that like the GAN output itself, such metrics are still ``twice removed'' from real data. 
We may have escaped comparing scientific data in terms of ``\%-catness'' but we still would make comparisons in terms of ``\%-simulation-domain-featureness'' which we then use as a measure of what is really ``\%-real-domain-featureness''.

Nonetheless, the LArTPC GAN-using community may come together to develop domain-specific xID-like metrics in this manner.
This would require strong coordination to standardize simulation configuration to generate training samples, the labels to use, what post-processing to apply, how to train Inception (or perhaps another network), distribution of pre-trained models and standardizing their application.
Different GANs target different data domains (raw ADC, signal processed, output of reconstruction) and thus some variety such metrics need construction.
Studies would also be needed to understand how these metrics differ when applied to data representing different LArTPC detector designs.

Another approach is to form a comparison metric by replacing the label-trained network with an auto-encoder (AE) network trained on domain-specific data samples.
The values of nodes forming the central bottleneck layer of the AE network can form the vectors used in a distance measure.
A major benefit of this approach is that it requires no labeling and thus can be trained on samples from all relevant domains (eg, real, simulated and ersatz).

Furthermore, some AE techniques allow for the latent vector space to be made smooth so that interpolating between different the points corresponding to two samples produces meaningfully similar decoded samples.
Ideally we may find some means by which we can understand a geometric metric tensor of this latent space so that we may form a distance with some semantic and quantitative meaning.
In section~\ref{sec:usage} we will suggest one way that may provide this.

\subsection{Invariant metrics}

When we apply GAN to scientific data we can exploit another validation strategy.
We may form metrics which are sensitive to certain conserved quantities, symmetries or correlations which are expected to be invariant over the features of any given sample. 
These are given to us by the physics of the detector operation and of fundamental particles and their interactions.

For example, raw waveforms read out from capacitive-coupled sense electrodes in LArTPC (``induction channels'') should integrate over long times to a value nearly zero (after baseline subtraction). 
Or, at high level after reconstructing GAN-generated data, particle momentum must be conserved and invariant mass peaks must be recovered at expected energies and Bragg peaks should appear with proper characteristics.
Where a GAN is involved in generating or where other networks may provide categorizing, we may validate their capability against these types of constraints.

Such metrics pose a conundrum. 
We may choose to bake them into the loss function in order to get a better trained network. 
But then, such a network will, by construction, produce results which are valid according to the metrics.

\section{Domain isomorphism}
\label{sec:ls4gan}

A novel approach to attacking the ``domain-equivalency mistake'' is being investigated as part of a lab-directed research and development project at Brookhaven National Lab. 
We are developing a method allowing a sequence of attacks on systematic errors.
First is to reduce the domain difference, then to quantitatively measure remaining difference and finally to provide a mechanism to propagate that remaining systematic difference through any subsequent processing and analysis.

The work goes under the name Large-Scale Scientific Simulation Systematic GAN (LS4GAN). 
It uses GAN-style networks to form a two-way isomorphic map that transforms samples (eg images) from one domain to another (eg ``real'' to ``simulated'' and vice versa). 
The network architecture and training is such that information specific to an individual sample is retained by the mapping while information that is general to the source domain is replaced by information general to the target domain.

For example, given a simulated detector readout of a neutrino interaction event, the features from the modeled noise and detector response would be replaced with equivalent features from the real detector. 
Meanwhile, the information which is specific to the particular neutrino interaction in the event will be retained. 
The simulation can be thought of as keeping its intended nature but being made ``more real''.

An existing proof of principle is demonstrated by CycleGAN~\cite{cyclegan2017},\cite{cyclegan2020}.
The LS4GAN effort has also produced an initial application of our variant network UVCGAN~\cite{uvcgan2022} on standard data photographic image sets.
The LS4GAN effort is now working to apply this technique to LArTPC data which confront novel challenges due to sample size and the fact we may not rely on the human bias of ``artistic judgment'' of the quality of results.

Once those challenges are overcome, we may exploit this mapping in a few ways. 
First, we may compare the output of the simulated-to-real mapping against its simulated input on a  sample-by-sample, pixel-by-pixel basis.
For example, a simple pixel-wise difference image may be formed and it is expected to highlight certain features which would otherwise be invisible in the conventional validation techniques such as described in section~\ref{sec:closer}. 
Given substantial domain expertise we may reasonably expect such high fidelity comparisons will point us to improvements we may make to the simulation and thus bring its subsequent output yet closer to reality.

After some number of iterations on the above process we expect to reach a point of diminishing returns. 
We may then retrain LS4GAN a final time and use the nominal simulation output and the more realistic simulated-to-real mapping to create two data sets of near identical sample pairs. 
The event-by-event difference holds a representation of whatever residual systemic difference exists between the simulated and the real data domains.

As these two datasets differ slightly in their information content but not at all in their format, each may be individually sent down a common reconstruction and analysis chain, be it based on conventional or AI/ML methods or some mixture. 
The results of processing each set may then be compared to form a quantitative estimate of the systematic uncertainty which due to our use of an imperfect simulation.

This will provide us a quantitative measure of the domain-equivalency mistake but it must be recognized it still neglects some residual uncertainty. 
The simulated-to-real mapping is itself imperfect and subject to systematic differences driven by, for example, the statistical sampling of simulated and real data that is made in defining a training set. 
So, the resulting systematic uncertainty is itself uncertain. 
The value of this residual unknown systematic uncertainly can arguably be estimated as being of some lower order compared to the systematic uncertainty measure the technique provides.
Arguing that becomes a detailed problem for the particular experiment applying this method.

\section{Trainable simulation}
\label{sec:diffsim}

A trained neural network encodes some learned information into an abstract representation over a number of parameters.
Any given parameter in that set has no identifiable meaning.
Here we describe techniques to develop models with semantically meaningful parameters that can be efficiently trained in the same manner as are neural networks. 
Two techniques are described and both provide computationally efficient methods to calculate the gradient over the learned parameters. 

\subsection{Backpropagation}

We may ``simply'' develop (or reimplement) a detailed simulation with code that uses a software library that provides the ``auto gradient'' feature. 
This feature allows the gradient of every function return value with respect to its arguments to be computed with no explicit code provided by the developer. 
Perhaps the most known examples of libraries providing this feature include PyTorch and JAX for Python while implementations for C++ applications also exist.
Given a means to compute gradients we may apply gradient descent optimizations such as are commonly used to train neural networks.
In fact, the resulting code is structured as a network of computation nodes in a very similar manner to that of neural networks differing only in terms of connectivity and functions.
When the auto gradient library provides support for GPUs, as many do, we gain the side benefit of hardware acceleration.

\subsection{Forward gradient}
Recent work~\cite{baydin2022gradients} has introduced a method to efficiently calculate an estimate of a gradient in the context of a forward pass of the simulation.
It is stochastic in nature which is perhaps particularly acceptable to calculate gradients through a simulation which is also stochastic.
In principle, this method may be applied to our ``usual'' C++ simulations running on CPU though this will likely prove prohibitively costly and still requires a reimplementation of our simulations to run on GPU.
More work is needed to see if this is a viable alternative to the auto-gradient based approach.

\subsection{Usage}
\label{sec:usage}

A broad class architectures have been collected under the general umbrella of ``simulation-based inference''~\cite{sbi}. 
In addition, a trainable simulation can provide a potentially powerful extension to the LS4GAN approach.
Instead of passing minted samples to LS4GAN we may directly integrate a trainable simulation network to the corresponding input nodes of the architecture.
Training will then optimize over the semantically meaningful parameters of the simulation network.
Resulting changes in these parameters can be interpreted as a measure of their systematic errors.

Such a scheme is potentially very powerful but needs to be tempered by the fact that the semantic parameters of the simulation represent a particular information basis that may not fully span the domain of the real data.
Certainly, this basis is more relevant than the one based on labels of cats, cars and cardigans of xID but a qualitatively similar domain-equivalency mistake will confound the results.
For example, optimizing against real data with sources of noise that are not modeled in the trainable simulation will cause systematic shifts to be attributed to the unrelated parameters.

We may also combine this conceptual thread with that of an AE-based metric in an attempt to provide some semantic mapping into the latent vector space.
For example, the gradient the latent vector with respect to the trainable simulation parameters may give semantic meaning to the latent space.
Metrics formed on that latent space can then be assigned semantically meaningful uncertainties.

\section{Summary}
\label{sec:summary}

We have highlighted a core systematic problem in applying analyses which utilize high-precision AI/ML networks that are trained on simulation and applied to real detector data which we name the ``domain-equivalency mistake''. 
We may reduce this mistake by applying data transforms based on domain expertise.
We suggest to develop more appropriate variants of existing metrics used by current GAN researchers as well as novel ones in order to provide more meaningful ways to compare the quality of a generated samples against those in a reference set.
A developing AI/ML-based method, LS4GAN, is described and expected to reduce, measure and propagate systematic differences between two data domains with high precision.
Finally, we describe trainable simulations which bring a promise to provide semantic meaning to measures of systematic error estimates based on training between the simulated and real domains.

\printbibliography

@article{mbsigproc1,
                  author = "MicroBooNE collaboration",
                  title = "{Ionization Electron Signal Processing in Single Phase LAr TPCs I: Algorithm Description and Quantitative Evaluation with MicroBooNE Simulation}",
                  eprint = "1802.08709",
                  archivePrefix = "arXiv",
                  primaryClass = "hep-ex",                  
                  journal = "JINST",
                  volume = "13",
                  pages = "P07006",
                  year = "2018"
                  }

@article{mbsigproc2,
                  autor = "MicroBooNE collaboration",
                  title= "{Ionization Electron Signal Processing in Single Phase LAr TPCs II: Data/Simulation Comparison and Performance in MicroBooNE}",
                  archivePrefix = "arXiv",
                  eprint = "1804.02583",
                  primaryClass = "hep-ex",
                  journal = "JINST",
                  volume = "13",
                  pages = "P07007",
                  year = "2018"
                  }

@inproceedings{cyclegan2017,
  title={Unpaired Image-to-Image Translation using Cycle-Consistent Adversarial Networks},
  author={Zhu, Jun-Yan and Park, Taesung and Isola, Phillip and Efros, Alexei A},
  booktitle={Computer Vision (ICCV), 2017 IEEE International Conference on},
  year={2017}
}

@misc{cyclegan2020,
      title={Unpaired Image-to-Image Translation using Cycle-Consistent Adversarial Networks}, 
      author={Jun-Yan Zhu and Taesung Park and Phillip Isola and Alexei A. Efros},
      year={2020},
      eprint={1703.10593},
      archivePrefix={arXiv},
      primaryClass={cs.CV}
}

@misc{baydin2022gradients,
      title={Gradients without Backpropagation}, 
      author={Atılım Güneş Baydin and Barak A. Pearlmutter and Don Syme and Frank Wood and Philip Torr},
      year={2022},
      eprint={2202.08587},
      archivePrefix={arXiv},
      primaryClass={cs.LG}
}

@misc{uvcgan2022,
      title={UVCGAN: UNet Vision Transformer cycle-consistent GAN for unpaired image-to-image translation}, 
      author={Dmitrii Torbunov and Yi Huang and Haiwang Yu and Jin Huang and Shinjae Yoo and Meifeng Lin and Brett Viren and Yihui Ren},
      year={2022},
      eprint={2203.02557},
      archivePrefix={arXiv},
      primaryClass={cs.CV}
}

@article{sbi,
author = {Kyle Cranmer  and Johann Brehmer  and Gilles Louppe },
title = {The frontier of simulation-based inference},
journal = {Proceedings of the National Academy of Sciences},
volume = {117},
number = {48},
pages = {30055-30062},
year = {2020},
doi = {10.1073/pnas.1912789117},
eprint = {https://www.pnas.org/doi/pdf/10.1073/pnas.1912789117}
}

@inproceedings{fid,
    author = {Heusel, Martin and Ramsauer, Hubert and Unterthiner, Thomas and Nessler, Bernhard and Hochreiter, Sepp},
    booktitle = {Advances in Neural Information Processing Systems},
    editor = {I. Guyon and U. V. Luxburg and S. Bengio and H. Wallach and R. Fergus and S. Vishwanathan and R. Garnett},
    pages = {},
    publisher = {Curran Associates, Inc.},
    title = {GANs Trained by a Two Time-Scale Update Rule Converge to a Local Nash Equilibrium},
    url = {https://proceedings.neurips.cc/paper/2017/file/8a1d694707eb0fefe65871369074926d-Paper.pdf},
    volume = {30},
    year = {2017}
}

@article{kid,
    title={Demystifying MMD GANs},
    author={Bi{\'n}kowski, Miko{\l}aj and Sutherland, Danica J and Arbel, Michael and Gretton, Arthur},
    journal={arXiv preprint arXiv:1801.01401},
    year={2018}
}

\end{document}